\documentclass[manuscript, nonacm]{acmart}

\usepackage{graphicx}
\usepackage{amsmath}
\usepackage{booktabs}
\usepackage{epigraph}
\usepackage{algorithm}
\usepackage{algpseudocode}
\usepackage{comment}
\usepackage{enumitem}
\setlist{nolistsep}
\usepackage{dblfloatfix}
\usepackage{subcaption}
\usepackage{array}
\usepackage{multirow}
\AtBeginDocument{%
  }

\begin{document}

\title{Deploying Speech-Driven 3D Facial Animation in Unreal Engine for Production-Ready Digital Humans}

\author{Alessandro Busacchi}
\affiliation{%
  \institution{Utrecht University}
  \city{Utrecht}
  \country{The Netherlands}}
\email{ale.busacchi@gmail.com}

\author{Kazi Injamamul Haque}
\affiliation{%
  \institution{Utrecht University}
  \city{Utrecht}
  \country{The Netherlands}}
\email{k.i.haque@uu.nl}

\author{Zerrin Yumak}
\affiliation{%
  \institution{Utrecht University}
  \city{Utrecht}
  \country{The Netherlands}}
\email{z.yumak@uu.nl}

\renewcommand{\shortauthors}{Busacchi, Haque and Yumak}

\begin{abstract}
Speech-driven 3D facial animation research has shown promising results, but most methods rely on representations that are not compatible with production pipelines. In this work, we present a deployable system that bridges this gap by enabling speech-driven 3D facial animation directly in Unreal Engine (UE) using ARKit-compatible representations. We construct 3DMEAD-ARKit dataset by converting the MEAD corpus into blendshape sequences using MediaPipe, and retrain FaceDiffuser and ProbTalk3D-X to generate stochastic and emotion controllable animations. We further develop a modular UE plugin with a Python backend that supports model selection, and parameter control. We compare the results to two existing commercial tools: Epic Games’ MetaHuman speech-driven animator and Nvidia Audio2Face with a perceptual user study. The results highlight the importance of comparisons among academic and commercial pipelines. We recommend watching the supplementary video. We also plan to do live demonstrations of our work at Siggraph 2026 conference.
\end{abstract}



\maketitle

\section{Introduction}
Speech-driven 3D facial animation has gained significant attention with the increasing use of digital humans. Recent deep learning approaches map speech to facial motion effectively, but most rely on datasets based on 4D scans \cite{FaceXHuBERT_Haque_ICMI23} or pseudo-3D reconstructions \cite{Probtalk3D_Wu_MIG24}, typically represented using non-semantic parametric head models. While efficient, these lack the semantic control and editability of animator-centric FACS-based (Facial Action Coding System) \cite{Ekman2002FACS} blendshape systems used in production. Instead, they adopt standardized representations such as- ARKit \cite{ARKIT} blendshapes for animating characters in engines like Unreal Engine (UE). However, the lack of large-scale, high-quality blendshape datasets creates a gap between research methods and production workflows.

We address these gaps by enabling speech-driven animation in a FACS-based blendshape space and presenting a deployable UE system. We construct 3DMEAD-ARKit dataset from MEAD \cite{kaisiyuan2020mead} using MediaPipe \cite{mediapipe}, retrain two of our published models-  FaceDiffuser \cite{FaceDiffuser_Stan_MIG2023} and ProbTalk3D-X \cite{ProbTalk3DX_Haque}, and develop a modular UE plugin for model selection, parameter control, and animation of production-ready characters, along with a perceptual evaluation framework.

\section{Related Work}
FACS represent facial expressions in terms of facial action units (AUs), which can be expressed as semantic blendshape coefficients. Several academic methods have adopted FACS representations for speech-driven facial animation models \cite{peng2023emotalk, cafetalk} but research models primarily focus on quantitative benchmarking \cite{WildWest_Haque} and do not provide frameworks for deployment. Commercially, NVIDIA Audio2Face \cite{nvidia_audio2face} and Epic's xADA \cite{xADA} demonstrate practical production integration through engine plugins and supporting tools. However, although they generate visually compelling animations, they are closed systems, limiting the integration of external models.

\begin{figure*}[t]
  \centering
  \includegraphics[width=0.95\textwidth]{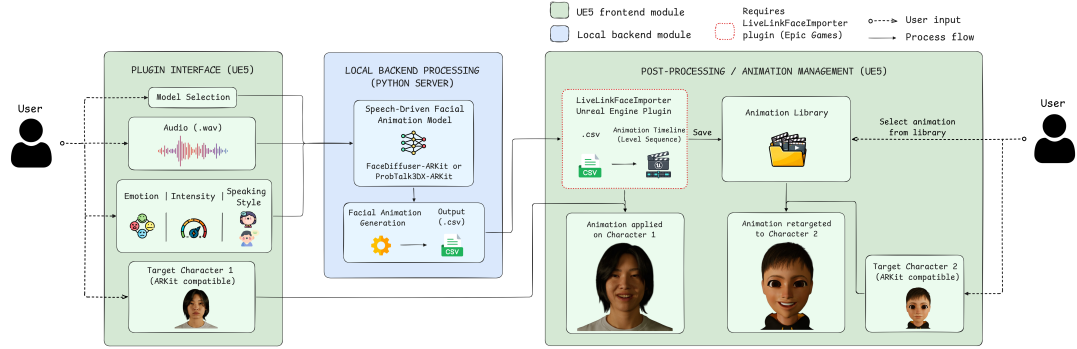}
  \caption{System overview: User can select the model, input audio (existing or live recording), conditioning style (i.e., speaking style, emotion, intensity) and a digital human character using the frontend interface. The data is passed to the backend for generating the animation data. Upon inference, the data is sent to the engine which is used to create, apply to selected character and save the animations in the animation library. User can also retarget the saved animations to other compatible characters.}
  \Description{The figure contains the system diagram showing the processes in action.}
  \label{fig:system}
\end{figure*}

\section{Methodology}
We construct the 3DMEAD-ARKit dataset from the MEAD corpus, which contains 47 speakers uttering 40 sentences across 8 basic emotions and 3 intensity levels, by extracting frame-wise blendshape sequences using MediaPipe \cite{mediapipe}. This dataset enables both emotion and intensity control when generating animation from speech. We retrain two of our previously published models- FaceDiffuser \cite{FaceDiffuser_Stan_MIG2023} and ProbTalk3D-X \cite{ProbTalk3DX_Haque}, with minor architectural modifications to train on the new dataset. More details on the model training can be found in the supplementary material. We then develop a UE plugin that deploys these models to animate ARKit-compatible characters from speech audio with controllable input parameters. We evaluate the system through expert evaluation with UE practitioners in Sec. \ref{sec:expert_eval}. Furthermore, perceptual user studies were conducted against commercial tools \cite{nvidia_audio2face, xADA} by employing statistically rigorous analysis presented in Sec. \ref{sec:perceptual_study}, unlike most research that conduct simple A vs. B user studies.

\paragraph{\textbf{Plugin Development}}
Our UE plugin bridges backend speech-driven animation models with the frontend production environment (see Fig. \ref{fig:system}). The interface allows users to select a model, character, input speech audio (pre-recorded or live), and control parameters such as speaking style, emotion and intensity. These inputs are sent to a local Python server for inference, which generates ARKit blendshape sequences and returns them as CSV data. The plugin leverages UE’s LiveLinkFaceImporter \cite{LiveLinkFaceImporter} to convert the data into animation sequences stored as \textit{Level Sequence} UE assets, enabling immediate playback and editing within the engine. The generated animation is automatically applied to the selected ARKit-compatible character. Additionally, an internal \textit{Animation Library} stores generated sequences as reusable assets, allowing users to retarget animations to other compatible characters in the scene.

\section{Practitioner Evaluation}
\label{sec:expert_eval}
We conducted two think-aloud practitioner evaluation sessions to assess the usability and user experience of the plugin. Participants were students from our university's \textit{Computer Animation} master course with prior experience in facial animation and UE. Each session lasted approximately 30 minutes, during which participants completed three tasks using the plugin. They successfully completed all tasks with minimal assistance and provided valuable feedback for future improvements. Additional details are provided in the supplementary material.

\section{Perceptual User Study}
\label{sec:perceptual_study}
We conduct two perceptual user studies evaluating \textit{Lip-Sync}, \textit{Realism}, and \textit{Expressiveness} on 7-point Likert scales following \cite{WildWest_Haque} for animations shown in randomized order, generated by four models: FaceDiffuser-ARKit (FD), ProbTalk3DX-ARKit (PT), NVIDIA Audio2Face (NV), and Epic Games' MetaHuman speech-driven animator (EG). Experiment 1 uses 12 test-set audios (6 male, 6 female) from dataset and includes an additional emotion recognition question against ground-truth emotion labels. Experiment 2 uses 8 in-the-wild audios (4 male, 4 female) generated with neutral emotion condition, and without emotion recognition question as no ground-truth labels are available for the in-the-wild audios. Surveys were hosted on Qualtrics survey tool, with participants recruited through Prolific (30+ per experiment who are mutually exclusive). More details about the perceptual user study experiments can be found in the supplementary material document. For data analysis, we perform within-subject repeated measures ANOVA (RMANOVA) to assess statistical significance across models for the three dependent variables (DVs) : \textit{Lip-Sync}, \textit{Realism}, and \textit{Expressiveness} where the independent variables (IVs) are the 4 models: FD, PT, NV and EG. 

[$H_0^L$ , $H_0^R$, $H_0^E$]: There is no difference in the mean scores across the four models (i.e., FD, PT, NV and EG) for [\textit{Lip-Sync}, \textit{Realism}, \textit{Expressiveness}].

\begin{figure*}[t]
  \centering
\begin{subfigure}{0.49\linewidth}
    \includegraphics[width=\linewidth]{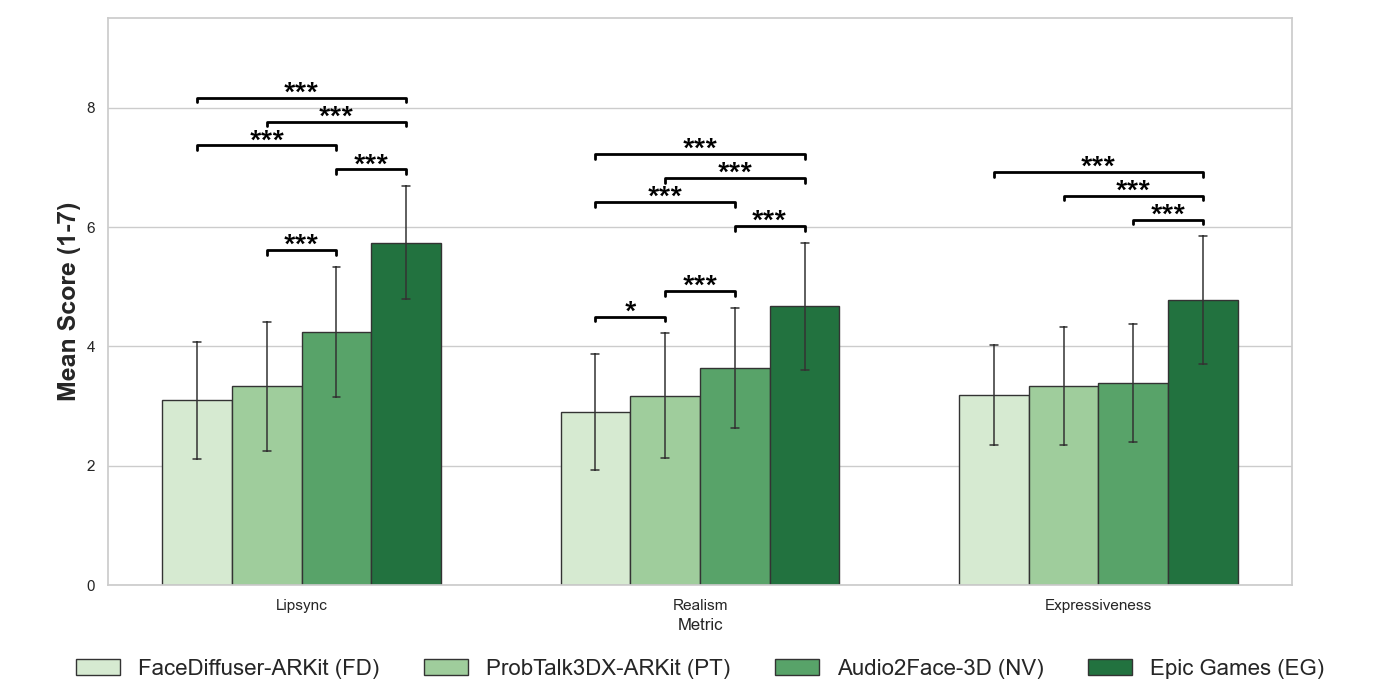}
    \caption{Experiment 1 Main Effect.}
    \label{fig:Experiment1}
  \end{subfigure}
  % \hfill
  \begin{subfigure}{0.49\linewidth}
    \includegraphics[width=\linewidth]{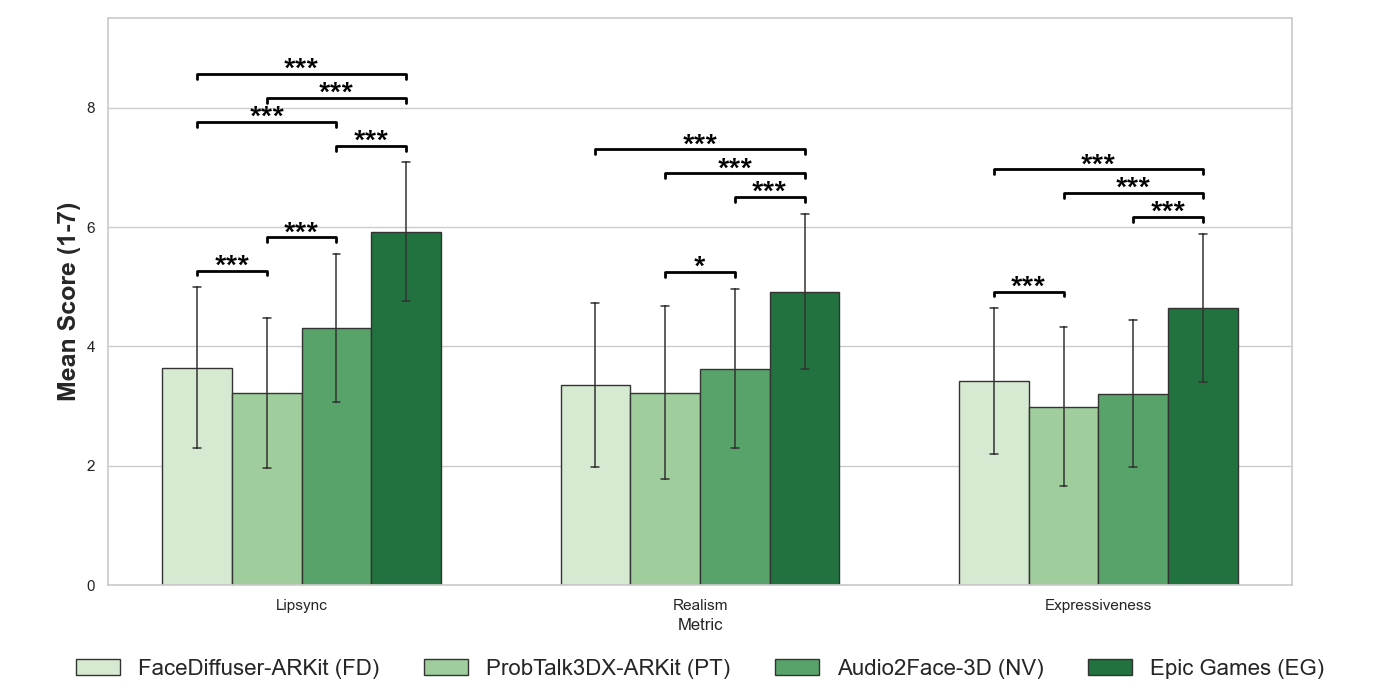}
    \caption{Experiment 2 Main Effect.}
    \label{fig:Experiment2}
  \end{subfigure}
  \caption{Experiment 1 and 2 results. Bar groups show mean ratings (error bars: SD) for Lip-Sync, Realism, and Expressiveness. Brackets with asterisks indicate significant Bonferroni-corrected pairwise comparisons (* $p<0.05$, *** $p<0.001$).}
  \label{fig:stat_fig}
  \Description{Figure containing main effect plots for experiment 1 and 2.}
\end{figure*}

\subsection{Experiment 1}

RMANOVA revealed a significant main effect for all perceptual metrics (Fig. \ref{fig:Experiment1}), rejecting $H_0^L$, $H_0^R$, and $H_0^E$. Significant effects were observed for Lip-Sync ($F(3,87)=139.20$, $p<0.001$, $\epsilon=0.63$, $\eta_g^2=0.51$), Realism ($F(3,87)=77.56$, $p<0.001$, $\epsilon=0.55$, $\eta_g^2=0.31$), and Expressiveness ($F(3,87)=70.10$, $p<0.001$, $\epsilon=0.72$, $\eta_g^2=0.31$). Post-hoc Bonferroni tests showed that EG significantly outperforms all methods across all perceptual dimensions ($p<0.001$). NV achieved the second-highest ratings in Lip-Sync and Realism, significantly outperforming FD and PT ($p<0.001$), while PT outperformed FD in Realism ($p<0.05$). No significant differences were found among FD, PT, and NV for Expressiveness. EG obtained the highest mean scores for Lip-Sync, Realism, and Expressiveness, followed by NV, PT, and FD. For emotion recognition, EG also achieved the highest accuracy (71.11\%), with NV (55.00\%), FD (51.11\%), and PT (49.72\%).

\subsection{Experiment 2}

For Experiment 2, RMANOVA revealed significant main effects for Lip-Sync ($F(3,87)=98.89$, $p<0.001$, $\epsilon=0.51$, $\eta_g^2=0.41$), Realism ($F(3,87)=47.81$, $p<0.001$, $\epsilon=0.51$, $\eta_g^2=0.20$), and Expressiveness ($F(3,87)=60.24$, $p<0.001$, $\epsilon=0.62$, $\eta_g^2=0.21$), consistent with Experiment 1 and rejecting $H_0^L$, $H_0^R$, and $H_0^E$ (Fig. \ref{fig:Experiment2}). Post-hoc analysis showed that EG significantly outperforms all methods across all perceptual dimensions ($p<0.001$). NV achieved the second-highest performance in Lip-Sync and Realism, significantly outperforming PT ($p<0.001$, $p<0.05$) and FD in Lip-Sync ($p<0.001$). FD outperformed PT in Lip-Sync and Expressiveness ($p<0.001$), while no significant differences were observed between FD and NV for Realism or Expressiveness. EG achieved the highest mean scores for Lip-Sync, Realism, and Expressiveness, followed by NV, FD, and PT.

\section{Discussion and Limitations}
The perceptual study shows significantly higher ratings for EG and NV, which we attribute to their training on high-quality proprietary datasets and model-specific optimization. EG leverages MetaHuman control rig specific capture pipelines for constructing datasets that are more expressive than ARKit, while NV is trained on subject-specific 4D scans adapted to solve for ARKit blendshapes. Although 3DMEAD-ARKit dataset allows to overcome some barriers regarding large-scale dataset, the quality of the dataset has limitations. MediaPipe enables scalable monocular video conversion to blendshape sequences but its processing introduces noise and temporal jitters. We expect that higher quality datasets and further model optimization would improve perceptual performance that match commercial systems. The subjective user study was conducted using high-end characters such as MetaHumans. We also expect the ARKit-based models to score better ratings on low-end cartoony characters. Our study highlights the lack of comparison between academic and industry work on 3D speech-driven facial animation and urge the research community to integrate such comparisons in their evaluations. In conclusion, we present a framework for integrating speech-driven animation models into production allowing further generation of user experiments.

\bibliographystyle{ACM-Reference-Format}
\bibliography{references}

\clearpage

\appendix

\section{Supplementary Material}
\subsection{3DMEAD-ARKit Dataset}
\begin{figure}[t]
    \centering
    \begin{subfigure}[b]{0.30\linewidth}
        \centering
        \includegraphics[width=\linewidth]{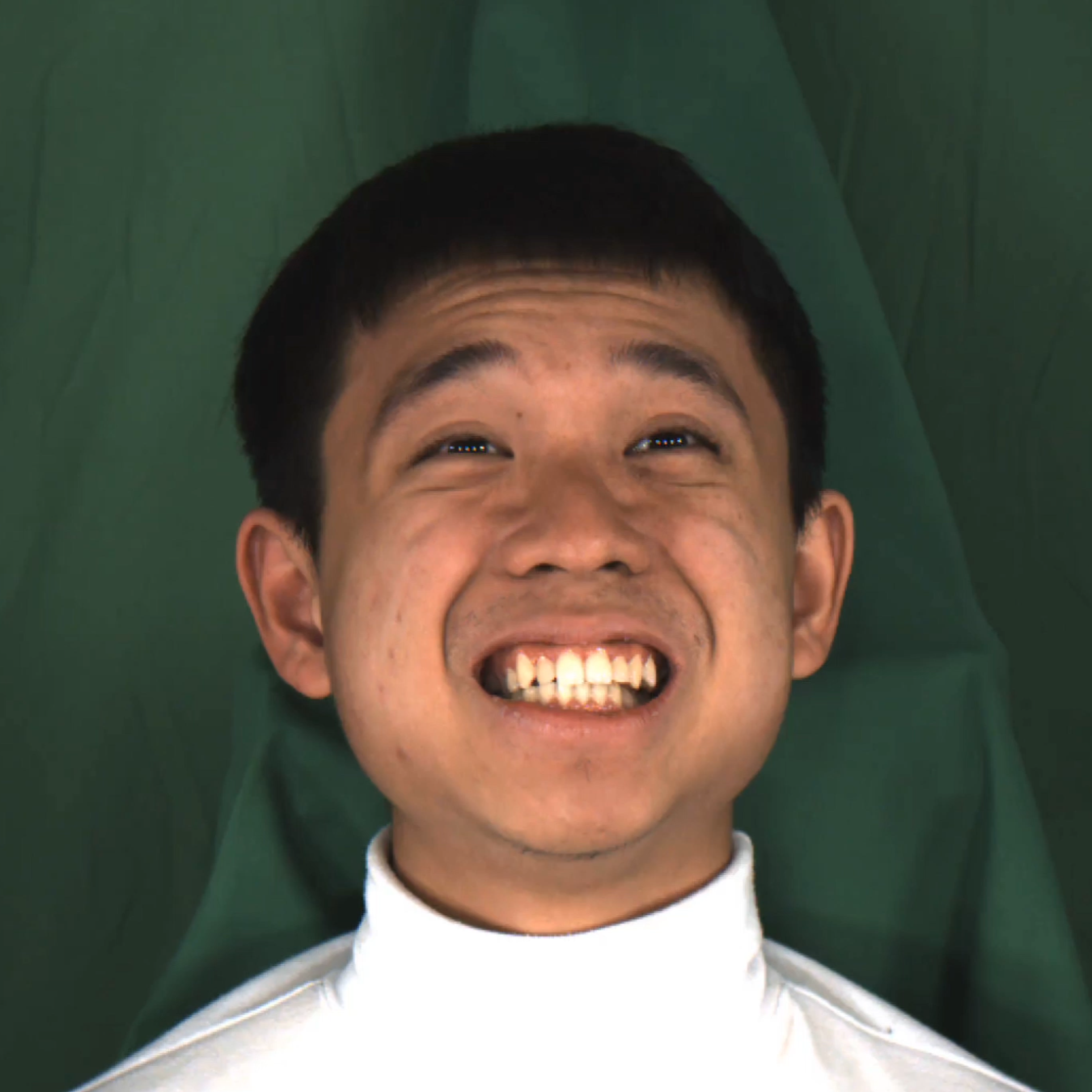}
        \caption{MEAD video frame}
        \label{fig:mead_frame}
    \end{subfigure}
    \hfill
    \begin{subfigure}[b]{0.30\linewidth}
        \centering
        \includegraphics[width=\linewidth]{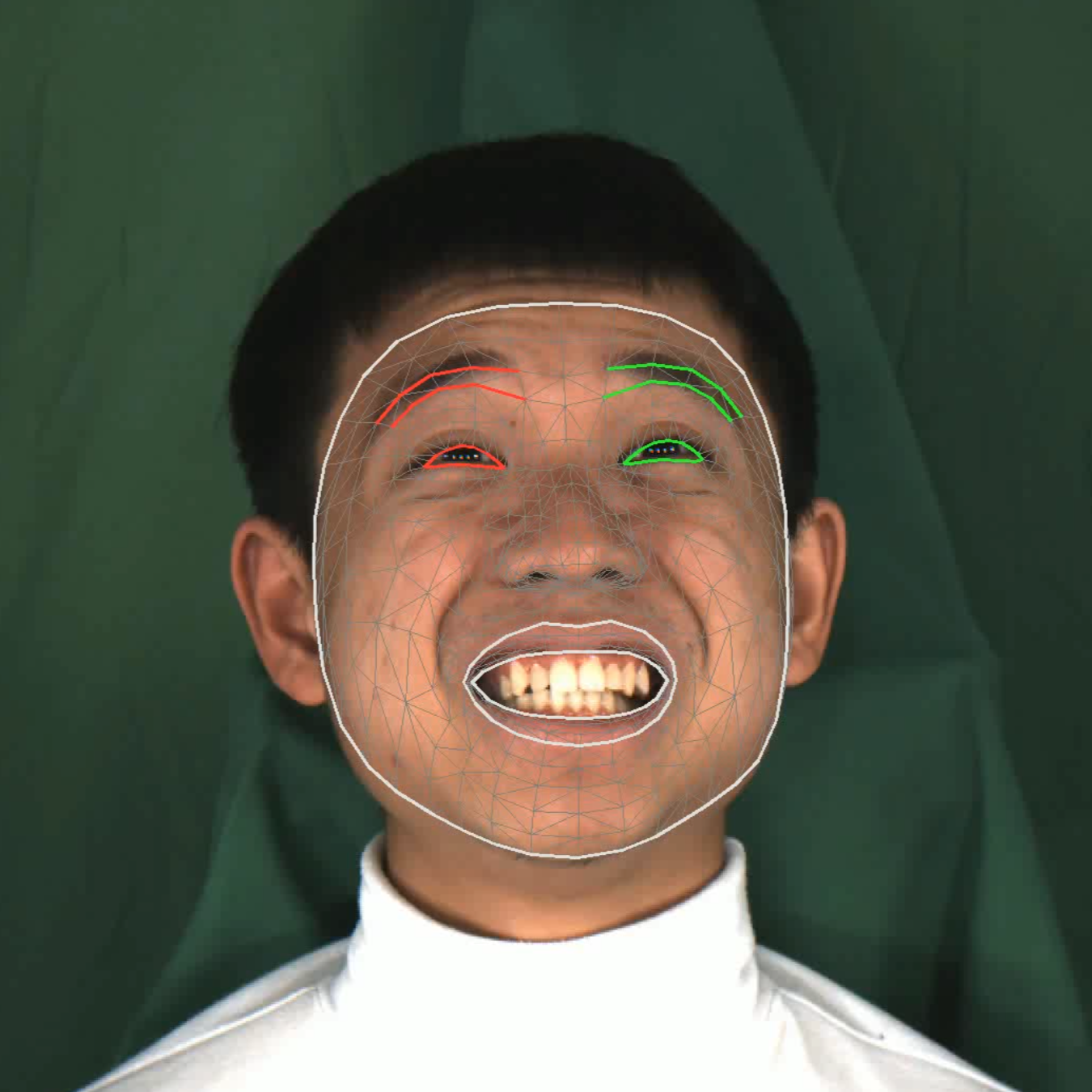}
        \caption{MediaPipe landmarks}
        \label{fig:mead_mediapipe}
    \end{subfigure}
    \hfill
    \begin{subfigure}[b]{0.30\linewidth}
        \centering
        \includegraphics[width=\linewidth]{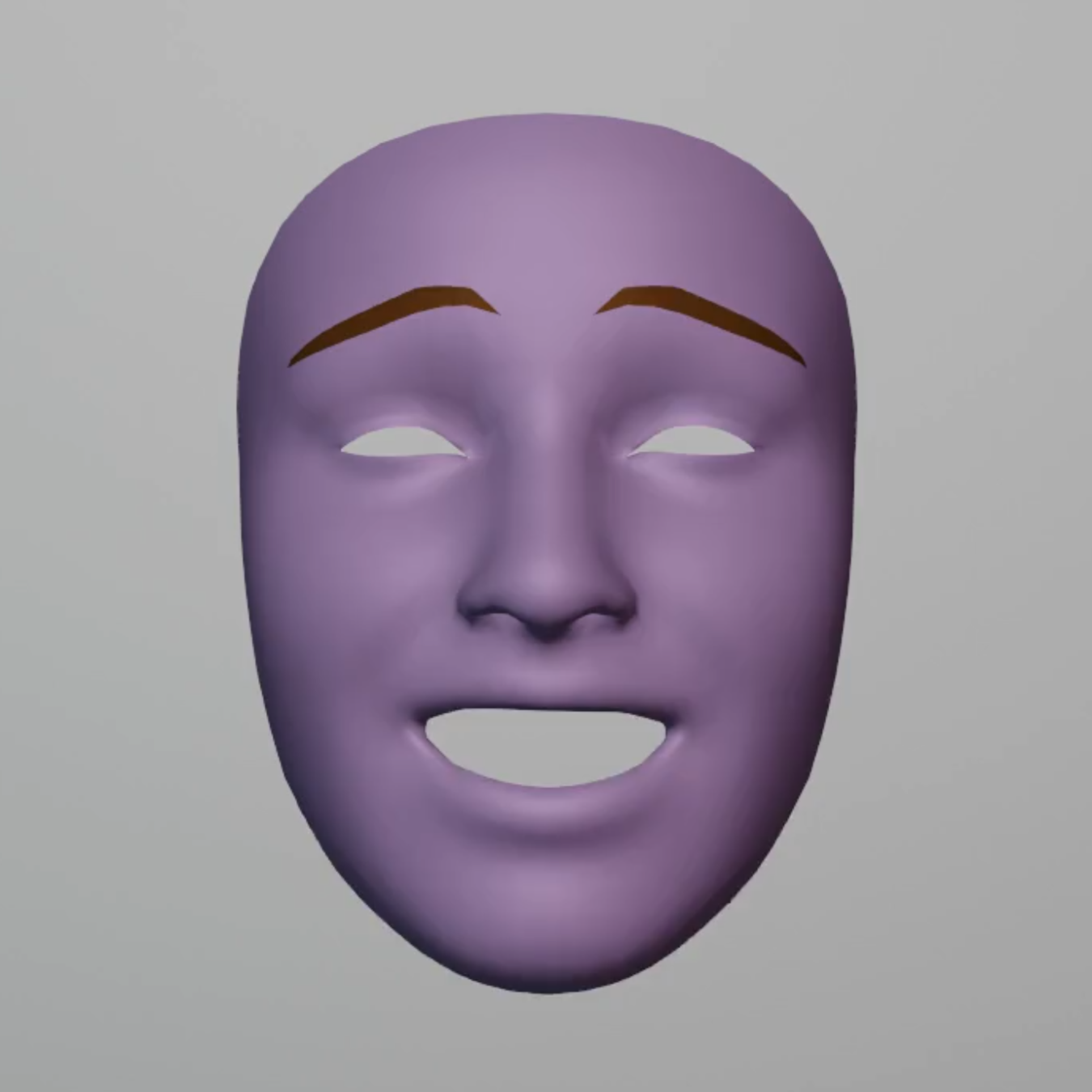}
        \caption{ARKit blendshape}
        \label{fig:beat_gt}
    \end{subfigure}
    \caption{Visualization of the dataset processing pipeline. From left to right: an original frame from the MEAD video dataset, the corresponding frame processed with MediaPipe showing detected facial landmarks, and the resulting ground-truth facial animation rendered on with an ARKit compatible face model using the extracted ARKit blendshape coefficients.}
    \label{fig:dataset_pipeline_example}
\end{figure}

To construct our 3DMEAD-ARKit dataset at scale, we utilize the large-scale MEAD video corpus \cite{kaisiyuan2020mead} and use MediaPipe \cite{mediapipe} to process the frame-wise video data into frame-wise ARKit blendshape coefficients. The RGB frames are passed to MediaPipe that first detects 3D facial landmarks and then the detected 3D landmarks are regressed into ARKit blendshape coefficients. Fig. \ref{fig:dataset_pipeline_example} demonstrates how an RGB frame (Fig. \ref{fig:mead_frame}) is used for detecting facial landmarks (Fig. \ref{fig:mead_mediapipe}) that are used to infer ARKit blendshape coefficients that can be used to animate ARKit compatible face (Fig \ref{fig:beat_gt}). In the constructed 3DMEAD-ARKit corpus of 47 subjects, reconstruction of ground-truth ARKit sequences using MediaPipe revealed visible noise and temporal jitters for several subjects upon closer inspection by rendering the full dataset. Consequently, the dataset was filtered to remove subjects with noisy reconstructed blendshape sequences, resulting in a final subset of 24 subjects used for retraining the models. 

\subsection{Training FaceDiffuser-ARKit and ProbTalk3DX-ARKit}
In order to retrain FaceDiffuser and ProbTalk3D-X on the newly constructed 3DMEAD-ARKit dataset, we conducted some experiments to find sufficiently working models for both that can generate qualitatively meaningful animations while respecting the input style condition (i.e., speaking style, emotion class and emotion intensity). Tab. \ref{tab:quantitative} demonstrates the quantitative evaluation of the retrained models using the metrics presented in \cite{WildWest_Haque}. Note that the metrics are computed in the ARKit blendshape space (i.e., each blendshape coefficient ranges from value 0 to 1) and not in vertex space. In Tab. \ref{tab:quantitative}, FaceDiffuser-ARKit and ProbTalk3DX-ARKit are the two final models selected for our plugin development. The quantitative evaluation results demonstrate that objectively, ProbTalk3DX-ARKit performs better than FaceDiffuser-ARKit. FaceDiffuser-ARKit model was slightly modified compared to the original B-FaceDiffuser model (which is the blendshape FaceDiffuser model version, trained on ARKit Data) where the style embedding is fused with the audio hidden representation before the latent is fed to the GRU decoder. Our experiments showed that the original architecture where the style embedding is fused after the GRU decoder, failed to generate emotion cues. It also incorporates additional weighted velocity and acceleration losses in training. For ProbTalk3DX-ARKit, the original model architecture was used with only changing the output dimension to match the output dimension of 3DMEAD-ARKit dataset and with a simpler loss for the reconstruction loss (as ARKit blendshapes are already normalized to [0,1]) unlike the original weighted reconstruction loss that is specific to the original model's FLAME-based 3DMEAD dataset. 

\begin{table}[t]
    \centering
    \resizebox{\textwidth}{!}{%
        \begin{tabular}{lcccccc}
            \toprule
            \textbf{Model} 
            & \textbf{MBE $\downarrow$ ($\times 10^{-1}$)} 
            & \textbf{LBE $\downarrow$ ($\times 10^{-1}$)} 
            & \textbf{MEE $\downarrow$ ($\times 10^{-1}$)} 
            & \textbf{CE $\downarrow$ ($\times 10^{-1}$)} 
            & \textbf{FDD $\downarrow$ ($\times 10^{-2}$)} 
            & \textbf{Diversity $\uparrow$ ($\times 10^{-1}$)} \\
            \midrule

            FaceDiffuser-ARKit 
            & 5.3783 & 3.8115 & 3.7742 & 3.6933 & 3.9647 & 0.8832 \\

            ProbTalk3DX-ARKit 
            & \textbf{5.0289} & \textbf{3.5639} & \textbf{3.3078} & \textbf{3.2269} & \textbf{1.6042} & \textbf{2.5384} \\
            \bottomrule
        \end{tabular}%
    }
    \caption{Quantitative evaluation of the two models trained on 3DMEAD-ARKit dataset. Best results are marked as \textbf{bold}. ProbTalk3DX-ARKit performs better than FaceDiffuser-ARKit in all quantitative metrics.}
    \label{tab:quantitative}
\end{table}

\subsection{Perceptual User Study Setup}

For both perceptual user studies (i.e., Experiment 1 and Experiment 2), we generated facial animation sequences using four speech-aniamtion generation models for comparison: FaceDiffuser-ARKit (FD), ProbTalk3DX-ARKit (PT), NVIDIA Audio2Face (NV), and Epic Games’ MetaHuman speech-driven animator (EG). All stimuli were rendered using MetaHuman characters to ensure a consistent comparison across methods, as EG only supports MetaHuman rigs. Specifically, we used the \textit{Aera} and \textit{Isaiah} MetaHuman presets for female and male characters respectively. The choice was made to select characters that are visually distinct from each other, while still maintaining consistency across conditions. The rendering setup employed a three-point lighting configuration (left, right, and center) to ensure clear facial visibility. A medium-dark green background was used to provide sufficient contrast with the characters.

We selected balanced male and female sequences in both experiments. Experiment 1 uses 12 test-set audio clips (6 male, 6 female), while Experiment 2 uses 8 in-the-wild clips (4 male, 4 female). In Experiment 1, different audio sequences were selected across conditions, using subject \textit{M013} for male and \textit{W018} for female sequences from the MEAD dataset while also balancing emotion category, and intensity level to reduce content-specific bias. Emotion intensities are evenly distributed across low, medium, and high levels. The audio clips were filtered based on duration and speech clarity to remove samples unsuitable for perceptual evaluation. In Experiment 1, we also evaluate the ability of each model to convey the intended emotion used for conditioning the facial animation generation, using labeled sequences from the dataset. Thereby we ensured that no contradictory emotion control condition was used (e.g., using emotion control condition as `Happy' while the audio represents a non-happy speech). FD and PT support discrete emotion categories with intensity control (low, medium, high). For NV and EG, which provide continuous parameter in [0,1] range, intensity values were mapped as 0.33 (low), 0.67 (medium), and 1.0 (high). However, not all emotion categories were included in the evaluation. We selected only those supported across all four models. While FD and PT support 8 emotions (neutral, happy, sad, anger, fear, disgust, surprise, contempt), \textit{contempt} was excluded as it is not supported by both NV and EG. Additionally, \textit{surprise} (EG) is represented as \textit{amazement} in NV; these correspond to different affective meanings and were therefore not treated as equivalent. In contrast, \textit{joy} (NV) and \textit{happiness} (FD/PT/EG) were treated as comparable affective categories and kept for evaluation.

\begin{figure}[t]
  \centering
  {\setlength{\fboxrule}{1.5pt}\setlength{\fboxsep}{0pt}%
    \fbox{\includegraphics[height=0.48\textheight]{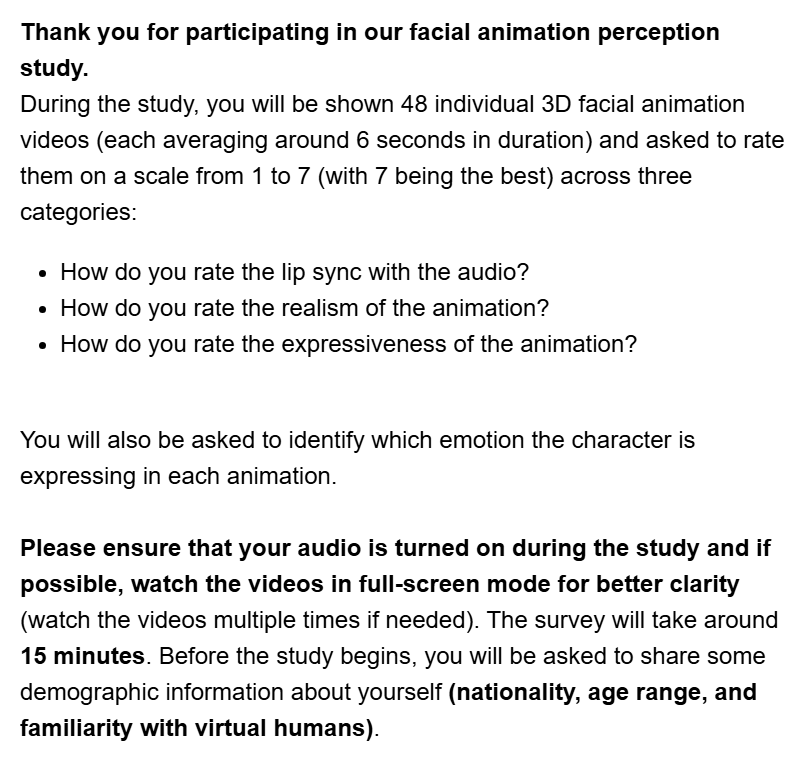}}}
    \caption{User study instruction.}
    \label{fig:user_study_consent}
  \Description{Figure containing screenshots of the user study UI 1.}
\end{figure}

\begin{figure}[htpb]
  \centering
    \includegraphics[height=0.95\textheight]{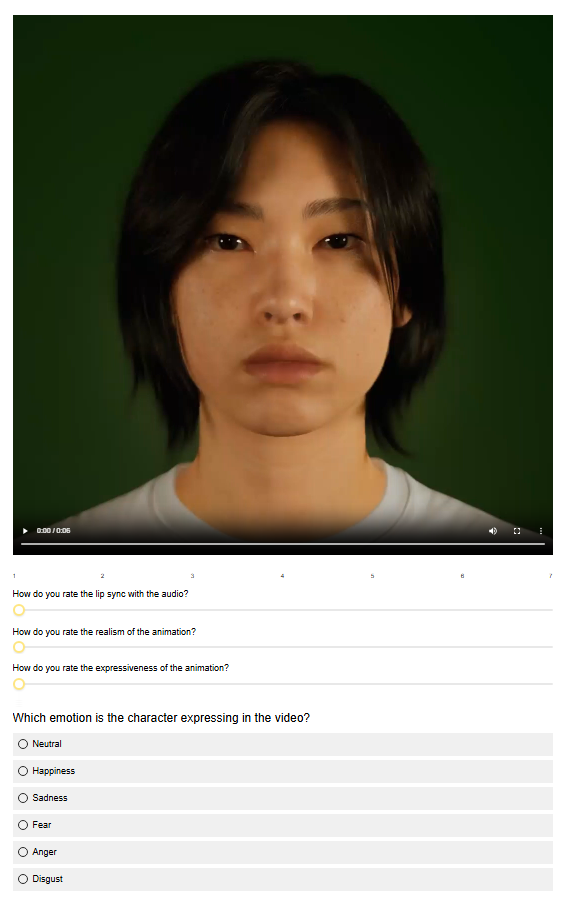}
    \caption{User study UI.}
    \label{fig:user_study_ui}
  \Description{Figure containing screenshots of the user study UI 2.}
\end{figure}

For Experiment 2, we selected in-the-wild movie audio clips, sourced from \url{https://movie-sounds.org/}, chosen based on plausibility with the target characters. As these clips are not annotated with emotion or intensity, the emotion recognition task was omitted and only perceptual ratings were collected. All models were conditioned on neutral emotion with low intensity for generation.

For inference with FD and PT, we used \textit{Man 1} and \textit{Woman 1} as speaking styles for male and female audio respectively. These correspond to speakers \textit{M003} and \textit{W009} from the MEAD dataset. For inference with NV, we selected the \textit{Claire} model for female audio and the \textit{James} model for male audio. Compared to \textit{Mark}, \textit{James} showed higher expressiveness in preliminary tests, consistent with the documentation (\url{https://docs.nvidia.com/ace/ace-unreal-plugin/2.5/ace-unreal-plugin-audio2face.html}). Additionally, we left \textit{Face Parameters} at their default values, as the documentation does not generally recommend modifying them. Since the other approaches do not provide post-processing parameters for animation control, we also kept all parameters at their default values to ensure a fair comparison. For inference with EG, we used the default decoder for generation, as the resulting outputs were already of high perceptual quality.

\subsection{Perceptual User Study Demographics}
In total, 35 responses were collected for Experiment 1, of which 5 were discarded due to failed attention checks, resulting in 30 valid responses. The age distribution was 23.33\% (18–24), 50\% (25–34), 3.33\% (35–44), 10\% (45–54), and 13.33\% (55–64). Prior to the study, participants rated their familiarity with virtual humans, 3D animated films, and video games on a 0–7 scale. The mean familiarity scores were 4.27, 5.97, and 6.33, respectively.

For Experiment 2, 33 responses were collected, with 3 excluded due to failed attention checks, resulting in 30 valid responses. The age distribution was 30\% (18–24), 43.33\% (25–34), 20\% (35–44), and 6.67\% (45–54). As in Experiment 1, participants reported their familiarity on a 0–7 scale prior to the perceptual task. The mean familiarity scores were 5.03 for virtual humans, 6.03 for 3D animated films, and 6.70 for video games.

The survey interface used for perceptual evaluation is shown in Fig. \ref{fig:user_study_ui} and Fig. \ref{fig:user_study_consent}. Furthermore, attention checks with intentionally mismatched audio–video pairs were embedded to identify and discard unreliable responses.

\subsection{Perceptual User Study Results Appendix}
\begin{figure}[t]
  \centering
\begin{subfigure}{0.49\linewidth}
    \includegraphics[width=\linewidth]{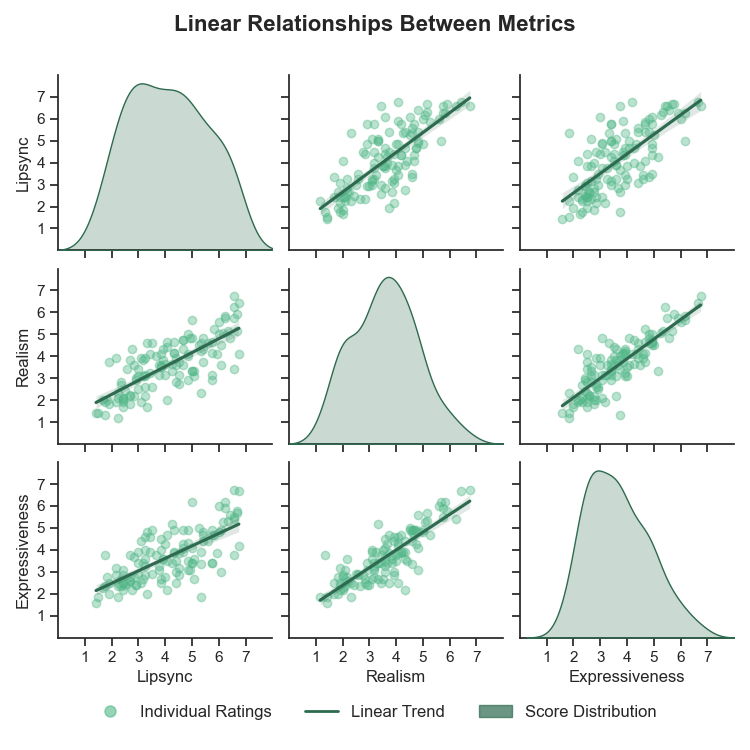}
    \caption{Experiment 1}
    \label{fig:Experiment1_corr}
  \end{subfigure}
  % \hfill
  \begin{subfigure}{0.49\linewidth}
    \includegraphics[width=\linewidth]{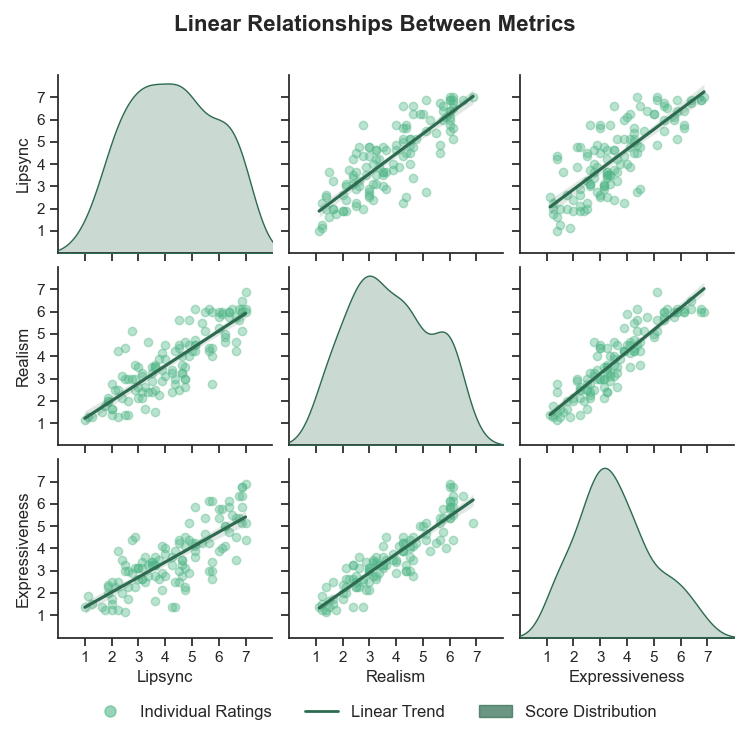}
    \caption{Experiment 2}
    \label{fig:Experiment2_corr}
  \end{subfigure}
  \caption{The plots demonstrate the correlation among the three perceptual metrics for Experiment 1 (\ref{fig:Experiment1_corr}) and Experiment 2 (\ref{fig:Experiment2_corr}).}
  \label{fig:stat_fig_corr}
  \Description{Figure containing correlations plots for experiment 1 and 2.}
\end{figure}

To further investigate the relationship between the perceptual metrics (i.e., Lip-Sync, Realism and Expressiveness), we conducted a two-way RMANOVA for both experiments. Results shows a significant interaction between models and metrics ($F(6, 174) = 26.71, p < 0.001$ for Experiment 1, $F(6, 174) = 22.04, p < 0.001$ for Experiment 2). This indicates that the gap between the models is not uniform across all the metrics (a model’s margin of superiority in Lip-Sync did not necessarily translate to an equal margin in Expressiveness). Also, repeated measures correlation ($r_{rm}$) analyses were performed to determine the correlation between the metrics. In Experiment 1, results showed exceptionally strong positive correlations across all the metrics (Lip-Sync and Realism ($r(89) = 0.93, p < 0.001$), Realism and Expressiveness ($r(89) = 0.88, p < 0.001$, and Lip-Sync and Expressiveness ($r(89) = 0.82, p < 0.001$). Experiment 2 led to similar pattern (Lip-Sync vs. Realism: $r = 0.88$; Realism vs. Expressiveness: $r = 0.89$; Lip-Sync vs. Expressiveness: $r = 0.82$), all significant at $p < 0.001$. These high coefficients suggest that the users perceive the three metrics as a unified indicator of overall quality. These relationships are summarized in Fig. \ref{fig:Experiment1_corr} and \ref{fig:Experiment2_corr}, for Experiment 1 and Experiment 2 respectively. The diagonal subplots display the Kernel Density Estimation (KDE) of the ratings, showing the concentration of scores and the overall performance distribution across the 7-point scale. The off-diagonal plots present the pairwise correlation between the metrics; the tight clustering of data points around the regression lines visually confirms the high $r_{rm}$ values found in our analysis. Also, the steep, positive slope of the linear trend lines across all pairs confirms perceptual dependency between pairs of dimensions. An increase in any single metric is associated with a proportional increase across the other perceptual metrics.

\subsection{Practitioner Evaluation Appendix}
\label{sec:expert_eval_appendix}

Two facial animation and UE practitioners were asked to complete three tasks: (i) generate an animation from a pre-recorded audio file using specified conditioning parameters, (ii) generate an animation using live microphone recording, and (iii) reapply a generated animation to a different ARKit-compatible character. Participants' reasoning was verbalized throughout the process while the researcher/developer intervened only when clarification was needed. Followed by a short semi-structured interview for additional qualitative feedback.

Both practitioners successfully completed all tasks with minimal assistance, indicating overall usability of the plugin. Feedback highlighted several strengths, including clear integration with native Unreal Engine workflows, practical usefulness for rapid animation prototyping, and effective support for previewing generated facial animations across multiple characters.

Several usability issues were identified. Both practitioners reported that the animation library and sequence re-application workflow lacked clarity, particularly due to some ambiguous terminologies and insufficient visual emphasis on key actions to be taken. The microphone workflow was considered unintuitive, with unnecessary interaction steps, unclear recording file management, and lack of automatic loading or preview of recorded audio. Additional suggested improvements included clearer indication of output file locations, actor-focused camera navigation after generation, improved audio preview support, and export options beyond UE-specific sequence assets to facilitate broader pipeline integration.

To summarize, the practitioners found the plugin promising for rapid facial animation iteration, while suggesting improvements primarily related to interface clarity, workflow streamlining, and export/generalization capabilities to other engines. The valuable UX feedback will be taken into account for our future iteration of the development cycle and software documentation.

\subsection{Asset Acknowledgment}
For our plugin demo, we utilize Epic Games' MetaHumans and a stylized cartoony model for visualization. The cartoony digital human is available on Fab with the name ``Cartoon Young Boy Rigged" made by giyas3dartist. The asset is available in the following link- \url{https://www.fab.com/listings/cb88681d-f0d3-4f6c-bfa6-ee38d2734f7f}. The scene and lighting setup used in both the plugin demo and perceptual user study were adapted from the ``MetaHuman Lighting" preset project available on Fab in the following link-  \url{https://www.fab.com/listings/52f008f2-bfd2-4db1-b9f5-94c5b1512b8a}.   

\subsection{Supplementary Video}
Supplementary video of our work can be viewed on our project webpage- \url{https://uuembodiedsocialai.github.io/AutoFaceARKit/}.

\end{document}